# Are Bushmeat Hunters Profit Maximizers or Simply Brigands of Opportunity?

Wayne Morra, Andrew Buck, and Gail Hearn

Wayne A. Morra
Arcadia University
Bioko Biodiversity Protection Program

Gail W. Hearn
Drexel University
Bioko Biodiversity Protection Program

Andrew J. Buck
Temple University

Working Paper
January 2011

## Abstract

Bushmeat hunters on Bioko Island, Equatorial Guinea use shotguns and snares to capture wild arboreal and ground animals for sale in the Malabo Bushmeat market. Two tools for the analysis of economic efficiency, the production possibilities frontier and isorevenue line, can be used to explain the post hoc changing spatial distribution of takeoff rates of bushmeat. This study analyzes changes in technical efficiencies over time and in different locations for the open access wildlife hunted on Bioko for the last ten years. Due to inadequate refrigeration in the field and the bushmeat market, animals must be sold quickly. The result is a takeoff distribution that is not efficient, consequently too many of the "wrong" species of animals are harvested. The larger, slower-breeding mammals (monkeys) disappear before the smaller, faster-breeding mammals (blue duikers and pouched rats), promoting a steepening of the production possibilities frontier, inducing a greater takeoff of monkeys than the expected efficient level. Soon after hunters penetrate into a new area, the relative selling price of monkeys exceeds the rate of transformation between ground animals and arboreal animals triggering inefficient and unsustainable harvests.

Contact information: morra@arcadia.edu, buck@temple.edu and gwh26@drexel.edu. Thanks to Conservation International, Margot Marsh Biodiversity Fund, Mobil Equatorial Guinea, Inc (MEGI), ExxonMobil Foundation, the Los Angeles Zoo, USAID, Marathon Oil and Hess Corporation for funding research expenses and in-country logistical support. Views expressed herein are those of the authors and do not necessarily reflect those of the MEGI, the LA Zoo, Hess Corp, Marathon Oil, the ExxonMobil Foundation or USAID. Thanks to Jose Manuel Esara Echube, Claudio Posa Bohome, Javier Garcia Francisco, Reginaldo Aguilar Biacho, Filemon Rioso Etingue and David Fernandez.

# Are Bushmeat Hunters Profit Maximizers or Simply Brigands of Opportunity?

**Introduction**

Bushmeat hunting on Bioko Island, Equatorial Guinea is an insignificant economic activity as a percent of GDP. Many of the (sub)species that are hunted on Bioko are endemic to Bioko (Fa, 1995) and their populations are hunted at unsustainable levels. As such, their extirpation from Bioko would constitute an irreversible loss to the world's biodiversity (Bergl, 2007). Rapidly rising income of the urban populace, due to vast petrochemical discoveries, is fueling demand for bushmeat. In the Malabo bushmeat market, meat sells for approximately $10/kilo, a delicacy even for the well-to-do Equatorial Guinean. This background begs consideration of the questions of economic efficiency and sustainable use of a natural resource.

Two tools for the analysis of economic efficiency, the production possibilities frontier and isorevenue line, can be used to analyze efficiency questions. These simple tools can explain the post hoc changing spatial distribution of takeoff rates of bushmeat. This study analyzes changes in technical efficiencies over time and in different locations for the open access wildlife hunted on Bioko for the last ten years. Due to inadequate refrigeration in the field and in the bushmeat market, animals must be sold quickly. The result is a takeoff distribution that is not allocatively efficient, consequently too many of the "wrong" species of animals are harvested. The larger, slower-breeding mammals (monkeys) disappear before the smaller, faster-breeding mammals (blue duikers and pouched rats), promoting a steepening of the production possibilities frontier, inducing a greater takeoff of monkeys than the expected efficient level. Soon after hunters penetrate into a new area, the relative selling price of monkeys exceeds the rate of transformation between ground animals and arboreal animals triggering inefficient and unsustainable harvests.

This paper: (1) estimates the technical, allocative, and access inefficiency of hunters in different geographic areas by constructing a production possibilities frontier and an isorevenue curve from the daily tallies of arboreal and ground animals hunted and sold in Malabo on Bioko Island during the last ten years; (2) documents changes in area-specific hunting intensity; (3) estimates the sustainability of commercial bushmeat hunting

**Natural and Political History of Bioko Island**

Bioko Island (2017 km$^2$) is a continental shelf island off the west coast of Africa. Primates are well represented on Bioko Island (Butynski & Koster 1996). Seven species of monkeys, arboreal animals for the purpose of this study, inhabit Bioko. Six of these seven species of primate are endemic subspecies, and because many are species now threatened throughout their continental ranges, Bioko Island is one of the world's "hotspots" for primate conservation (Cowlishaw, 1999). Only two species of hoofed mammals, both forest antelope, remain on Bioko Island. A number of other mammals are also large enough to hunt. For example, the brush-tailed porcupine and giant pouched rat (known locally as 'ground beef').

Efficient hunting methods considerably reduced Fernando Po's wildlife by the time of Equatorial Guinea's independence from Spain in 1968, at which time the Island was renamed "Bioko". The changes precipitated by independence included a general ban on firearms, collapse of Bioko's cocoa, coffee and cattle industries, and a greatly reduced human population. An estimated third of the population of 400,000 of Equatorial Guinea were either killed or fled into exile (UNHCR, 2001). The ban on firearms and reduction in human population favored forest regeneration and wildlife and, as a result, forest mammal populations began to recover during the 1970s and 1980s (Butynski and Koster, 1994).

The recovery of wildlife was short-lived. A commercial bushmeat market appeared in Malabo during the early 1980s and hunting to supply animals for this market became increasingly more organized during the 1990s. Since the mid-1990s, three factors have combined to place intense hunting pressure on the remaining populations of large forest mammals. First, as a result of the development of offshore oil extraction, local people have more money for bushmeat. Higher incomes have increased demand, driving the prices higher and making commercial hunting more profitable. Second, the larger mammals favored by consumers and hunters typically have long periods to sexual maturity and a slow reproductive rate, resulting in a slow population growth rate, making them vulnerable to over-harvesting. And third, as hunters enter the most remote parts of Bioko, they are now aided by newly paved roads from Malabo to the outlying towns. The new roads also make Bioko's two "protected" areas, Pico Basilé National Park and Gran Caldera & Southern Highlands Scientific Reserve accessible to the hunters.

**Method of Data Gathering**

The data upon which the conclusions of this paper are based come from two sources, both collected by the Bioko Biodiversity Protection Program (BBPP). First, a trained census taker records the animals arriving for sale at the only bushmeat market in Malabo from 08:00 - 12:30 six mornings/week. During a 10 year period, carcasses at the Malabo market were counted on 2,869 mornings (or market days: mean market days/month = 24.1, s.d. = 3.5) involving 113,174 carcasses. Imported bushmeat was excluded. Recorded data included species, age (adult or immature), sex, condition (alive, fresh, smoked), method of capture (snare or shotgun), the carcass' point of origin, and selling price. At various times from February 2002 through November 2007, weights and physical size measurements of bushmeat species have been obtained.

**Brief Overview of the Malabo Bushmeat Market**

Twenty-three species of animals from Bioko are available for sale, with varying degrees of regularity, at the Malabo bushmeat market. The IUCN Red List Categories (IUCN, 2007) highlight the threat of the bushmeat trade to Bioko's monkeys with all seven of the species either classified as 'Endangered' or 'Critically Endangered.' Over the last ten years, the most common animals sold in the Malabo bushmeat market in terms of biomass are blue duiker (31%), monkeys (26%), red duiker (18%), porcupine (10%), pouched rat (6%), python (4%) and monitor lizard (3%).

Figure 1 shows the annual mean biomass of ground and arboreal carcasses/day sold at the Malabo bushmeat market from January 1998 – December 2007, inclusive.

Figure 1. Mean biomass of ground and arboreal carcasses/market day by year at the Malabo bushmeat market, Bioko Island (January 1998 – December 2007, n = 113,174 carcasses, imports excluded).

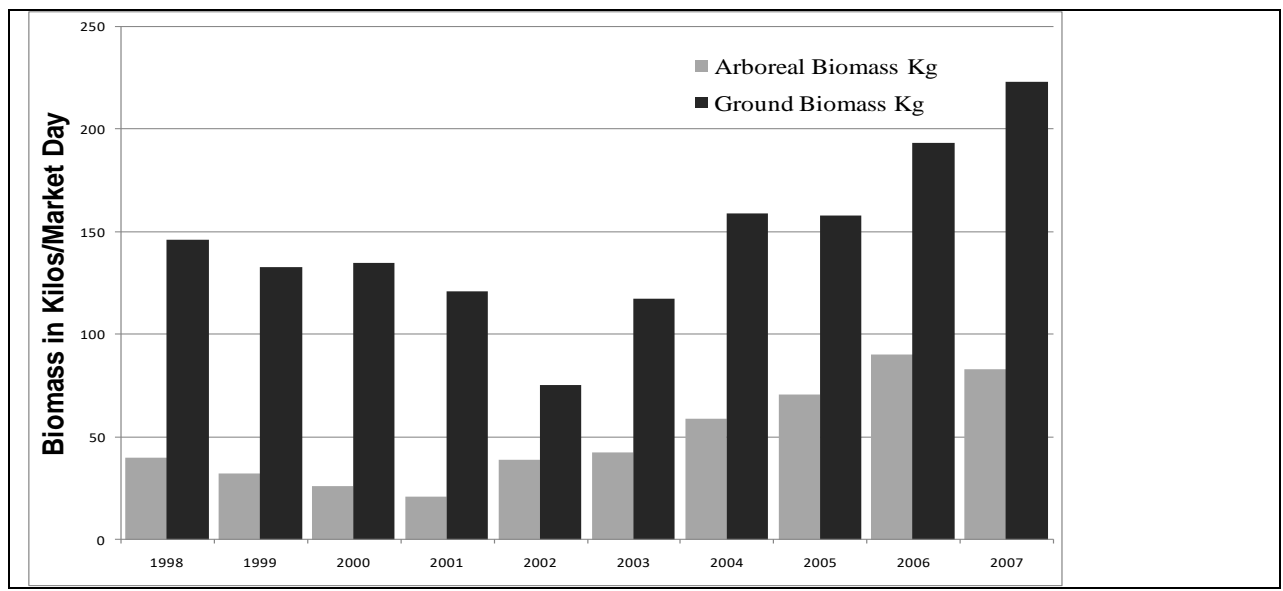

The Malabo bushmeat market has grown both in the number of carcasses and in revenue over the 120 months since January 1998. Using the U.S. Central Intelligence Agency (2007) estimate of 5% inflation per year, real average revenue from recorded sales increased 302% since 1998. The price of the largest monkey, the drill, increased by 257% during the 120 months covered by this study. The smallest increase in price, 85%, was for the Black Colobus. The increase in real sales must then reflect an increase in not only price, but also carcasses.

One hundred and forty-two hunters/trappers in 21 locations on Bioko were surveyed in 2003. The survey included all the significant hunting camps on Bioko. A "hunter" is defined as any person who spends at least part of his time hunting with a gun, even though many of them also use traps; "trappers" are those who only use traps. The median time in a hunting camp is 5.2 years, far below the mean of 13.9 years, indicating that a high proportion of the respondents are recent arrivals at their current location. Shotgun hunting is the only significant threat to Bioko's monkeys, accounting for 99% of the monkey kills. The pouch rat, porcupine and pangolin are largely harvested using traps. Other species, like the blue and Ogilby's duiker are increasingly hunted with shotguns.

Whereas the owner of a renewable resource takes into account the effects of resource depletion, the hunter (non-owner) of an open access renewable resource does not. Since the individual hunter does not include the cost of the decreasing availability in his optimal foraging calculation, the hunter, even if he is a rational calculator, will over-utilize an open access resource.

Aggravating the situation is the fact that bushmeat is not a single homogeneous resource. Because species grow, reach sexual maturity and reproduce at different rates, some popular bushmeat species (blue duiker) are still relatively common on Bioko, while others (Ogilby's duikers and monkeys) are increasingly rare. Hunters shoot anything profitable without regard for rarity; taking the rare species without regard for depletion of the common pool.

Figure 2 shows the declining percentage of animals harvested from the northern half of Bioko, an area that is readily accessible from Malabo. The percentage gathered from the northern half of Bioko does not decline monotonically. During 2003 a road was graded for a water project. The new road allowed access to a previously unexploited area on the western slope of Pico Basilé. Hunters moved in and over the next 3 years quickly hunted out most of the larger monkeys.

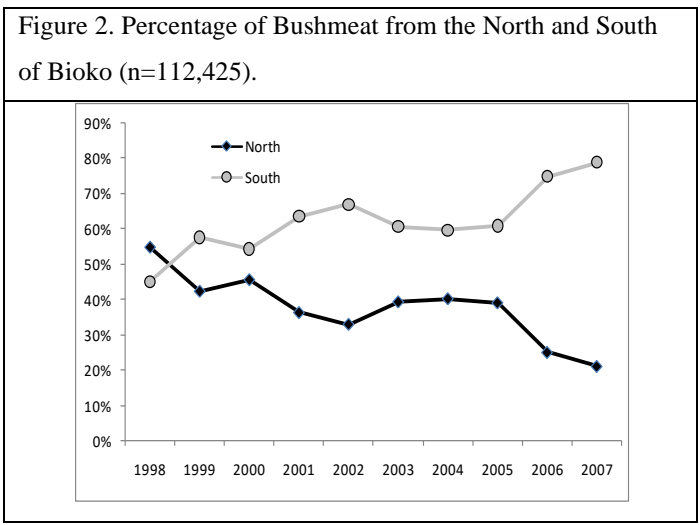

Figure 2. Percentage of Bushmeat from the North and South of Bioko (n=112,425).

Signs of hunting (e.g., spent shotgun shells, new hunting camps, increased encounters with hunters along census trails) and reduced rates of encounter with monkeys, duikers and other hunted species, indicate that there is increased hunting throughout the southern half of Bioko. During 2004, hunters began entering the Gran Caldera de Luba (19 km$^2$), a remote and nominally protected area, with greater regularity. Until 2004 the Caldera had been almost completely free of hunting. In 2004, after a cessation of funding, the BBPP's passive guarding/monitoring program was temporarily suspended. Within months hunters quickly seized the opportunity and began hunting in the Caldera, thus explaining the rising percent coming from the southern end of the island.

**Methodology**

Let the joint production function be

$$e^{(\alpha-a^2 s)t} e^{[\gamma-g^{1.9}(L-s)t]} = 0 \tag{1}$$

Where $\alpha$ is the instantaneous rate of growth of the arboreal animal population, $\gamma$ is the instantaneous rate of growth of the ground animal population, a is the rate at which arboreal animals are being harvested, and g is the rate at which ground animals are being harvested. L=1 is the total amount of labor input, and s is the proportion spent in shotgun hunting. An increase in s can be interpreted as meaning either that hunters are better at the shooting or they have found a location where the arboreal animals are more abundant. The function is strictly quasi-convex in the harvest rates and corresponds to the PPF of Figure 3.

Figure 3. Hunters' Production Possibilities Frontier.

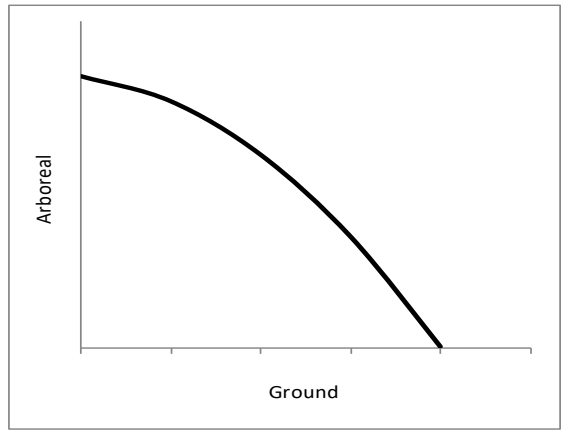

We can also use this function to describe the two populations over time using some parametric assumptions: L = 1, $\alpha$ = .02, a = .019, s = .5, $\gamma$ = .04 and g = .02 (see Figure 4). By increasing the harvest rates we can slow population growth and even cause it to decline. As the parameters a or g increase (Figure 4) the arboreal and ground population paths become flatter. As $\alpha$ or $\gamma$ increase the paths become steeper and more convex.

Figure 4. Animal population growth and abundance.

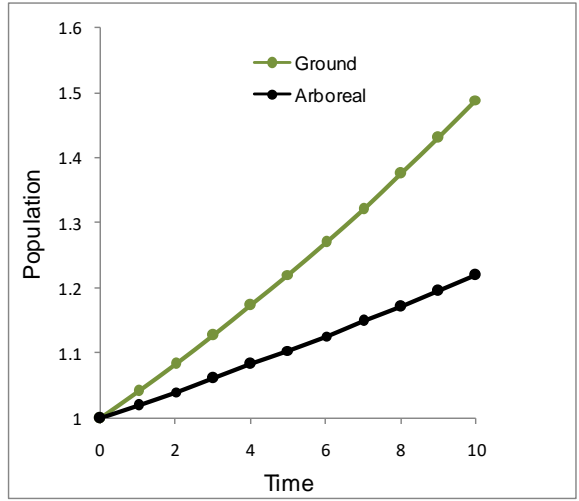

Using the implicit function theorem we can find the rate of transformation between ground and arboreal animals; the slope of the production possibilities frontier (see Figure 5).

$$\frac{dg}{da} = \frac{-\dfrac{d}{da}\left[e^{(\alpha-a^2 s)t} e^{(\gamma - g^{1.9}(L-s))t}\right]}{\dfrac{d}{dg}\left[e^{(\alpha-a^2 s)t} e^{(\gamma - g^{1.9}(L-s))t}\right]} = \frac{-1.053as}{g^{.9}(1-s)} \qquad (2)$$

The slope is negative as guaranteed by the assumption of quasi-convexity of the joint production function. As more time is needed to hunt monkeys relative to other animals, the PPF becomes steeper. This is illustrated in Figure 5.

Figure 5. Hunter Inefficiency and the slope of the PPF.

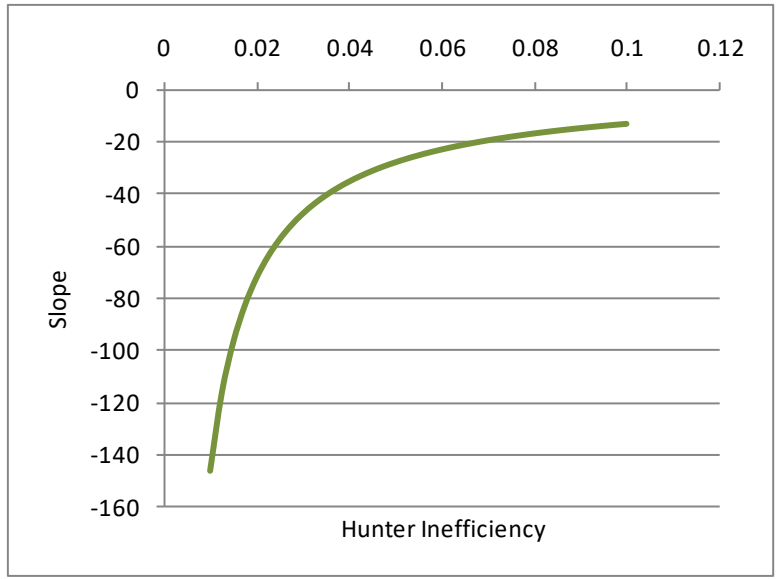

With some more parametric assumptions we can show the same thing by plotting two PPF's. The flatter PPF (Figure 6) is when hunters become less effective in hunting monkeys, represented by an increase in s and a flatter PPF.

Figure 6. Hunter Inefficiency and the PPF.

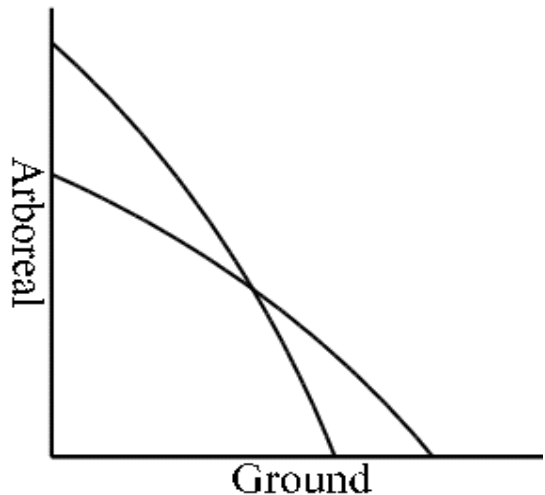

Animals in the forest are a renewable open access resource, as such they can be subjected to excessive hunting pressure. Therefore under proper hunting management, there are sustainable harvest rates of ground and arboreal animals. A sustainable harvest rate is that which does not result in collapse of the population of the target species. We take sustainable to mean no declining populations. A myopic harvest rate results in population collapse and is typical of open access resources. The difference between sustainable and myopic harvest rates is shown in Figure 7.

Figure 7. Sustainable and Myopic PPF's.

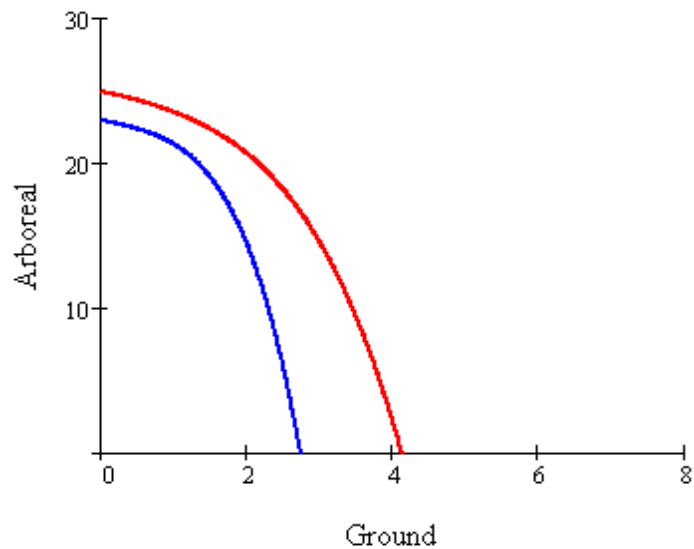

The red PPF shows current practice, which disregards the question of sustainability. Even though sustainability is not part of the red PPF, those points are "efficient" in the sense that all resources dedicated to hunting are fully employed. The red PPF is a short run, myopically efficient set of hunting combinations.

The blue PPF shows the sustainable harvest combinations, given the stock of animals, stock of habitat, labor inputs, and hunting technology/knowledge. A point on the blue PPF is a sustainably, productively efficient combination of ground and arboreal animals harvested. The blue PPF is a long run curve, meaning that the myopia problem has been solved.

A combination inside the blue curve is a sustainable combination, but it is not long run efficient. A harvest combination between the two curves is a myopic choice, since that harvest rate is not sustainable. In addition, it is not even efficient in the myopic short run.

Figure 8. Productive and Allocative Efficiency.

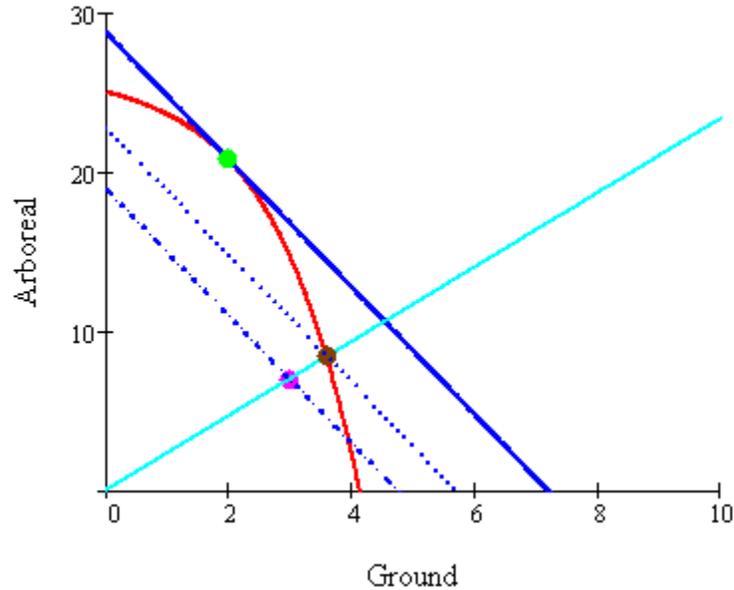

In Figure 8 only the myopic PPF is shown. The downward sloping solid blue line is the iso-revenue curve for the hunting industry. Given resources, prices and technology, the efficient choice is at the green dot. However, suppose that hunters have harvested the combination at the magenta dot. How do we measure the inefficiency of choosing the magenta dot instead of the green dot? Given the magenta harvest, if hunters were to increase production of ground and arboreal animals at a constant proportion then they would expand along the cyan ray from the origin to the black point. Any point along the red PPF below the black point would generate less revenue. Any point on the red curve above the black point would generate more revenue, but it would also move hunters toward the optimal choice of ground and arboreal animals. Therefore choosing a point above the black harvest involves eliminating unemployment of resources and some part of the loss due to changing the allocation between ground and arboreal animals. For our purposes we don't want to mix the two sources of inefficiency.

Note that the choice of the black point is not the shortest distance from the magenta choice to the PPF. The shortest distance point would be the least squares projection of the magenta point onto a line tangent to the red PPF. This point would lie below the black point, resulting in less revenue to the hunter and would be less desirable from the hunters' perspective.

A simple numerical example aides in the interpretation of Figure 8. In the figure the price of a ground animal is 4 and the price of an arboreal animal is 1. The revenue generated from the magenta harvest is 4*3+7 = 19. The revenue generated from the black harvest is 4*3.6043+8.41 = 22.827. The revenue generated from the green harvest is 4*2 + 20.8 = 32.8. From these revenues the value of inefficiency due to unemployed resources, or technical inefficiency is 3.827. The loss in value due to the misallocation between ground and arboreal animals is 32.8 - 22.827= 9.973.

In Figure 9 we take up the inefficiency due to hunting at an unsustainable rate.

Figure 9. Efficiency and Unsustainable Hunting.

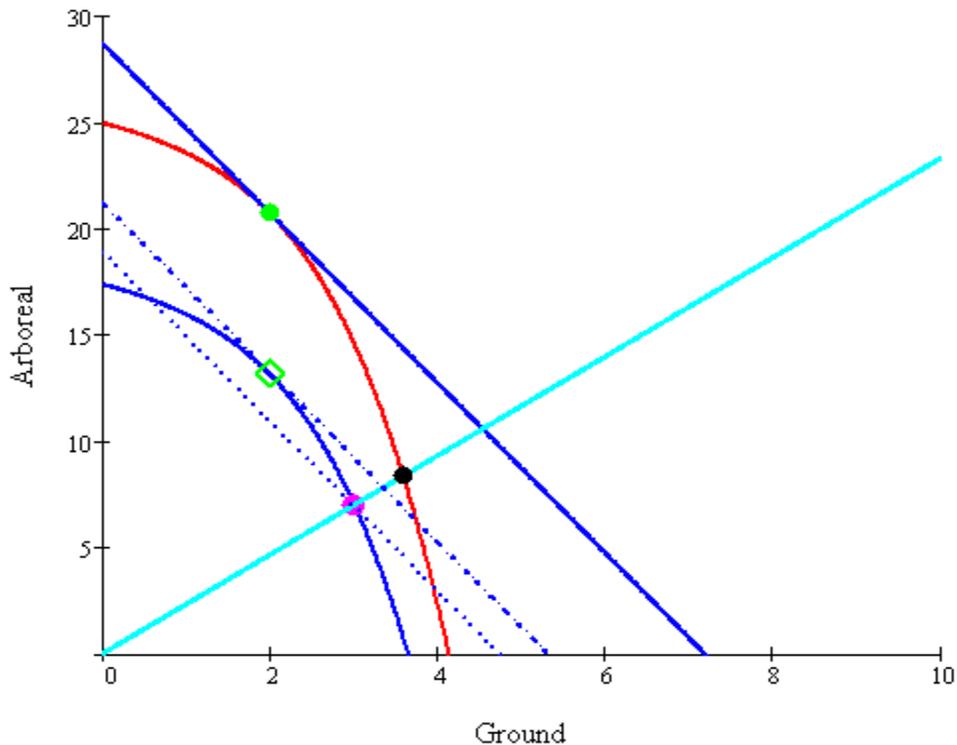

In this figure hunters again bring in the harvest at the magenta point. This is a sustainable harvest, but it is not allocatively efficient. Given prices, animal stocks, habitat and technology, they should have brought in the combination at the green diamond. The green diamond would result in greater sustainable revenue for the hunters. The green diamond results in a long run gain over either the myopic productively efficient black combination or the optimal myopic green dot combination. We can calculate the actual gains and losses from the data collected at the Malabo bushmeat market. There are three major hunting areas on Bioko Island: the northern half dominated by Pico Basilé, a 3,011m extinct volcano, but excluding the lowlands. The second region is the southeastern quarter where Riaba is the largest village. The final region is in the southwestern quadrant centered on the town of Luba. All three areas transport the majority of their catch to the capital city, Malabo, to obtain the high prices the urban market affords. We examine the technical and allocative efficiency of hunters for the years 1999, 2001, 2003, 2005 and 2007 in the three areas.

Daily biomass for 2,709 days were separated by origin of capture (Pico Basilé 643 days, Riaba 974 days and Luba 1,092 days) to construct the production possibilities frontiers for the different years. The allocatively efficient combination of ground and arboreal takeoff was determined by the tangency of the PPF and the isorevenue curve with slope equal to the weighted price ratio of the price of ground animals to the price of arboreal ($P_G/P_A$). Figure 14 illustrates the procedure for Riaba using 2007 data. Bushmeat arrived from Riaba and was counted at the Malabo bushmeat market on 259 days, represented by the black circular data points. The outer boundary of the data points constitutes the production possibilities frontier. A ray from the origin was extended through each point to the boundary, establishing the technically efficient combination of ground and arboreal takeoff. The distance from the observed data point to the technically efficient point is the amount of technical inefficiency. The distance from the technically efficient point to the tangency of the isorevenue line and the PPF is allocative inefficiency.

Figure 14. Technical and Allocative Efficiency in the Field.

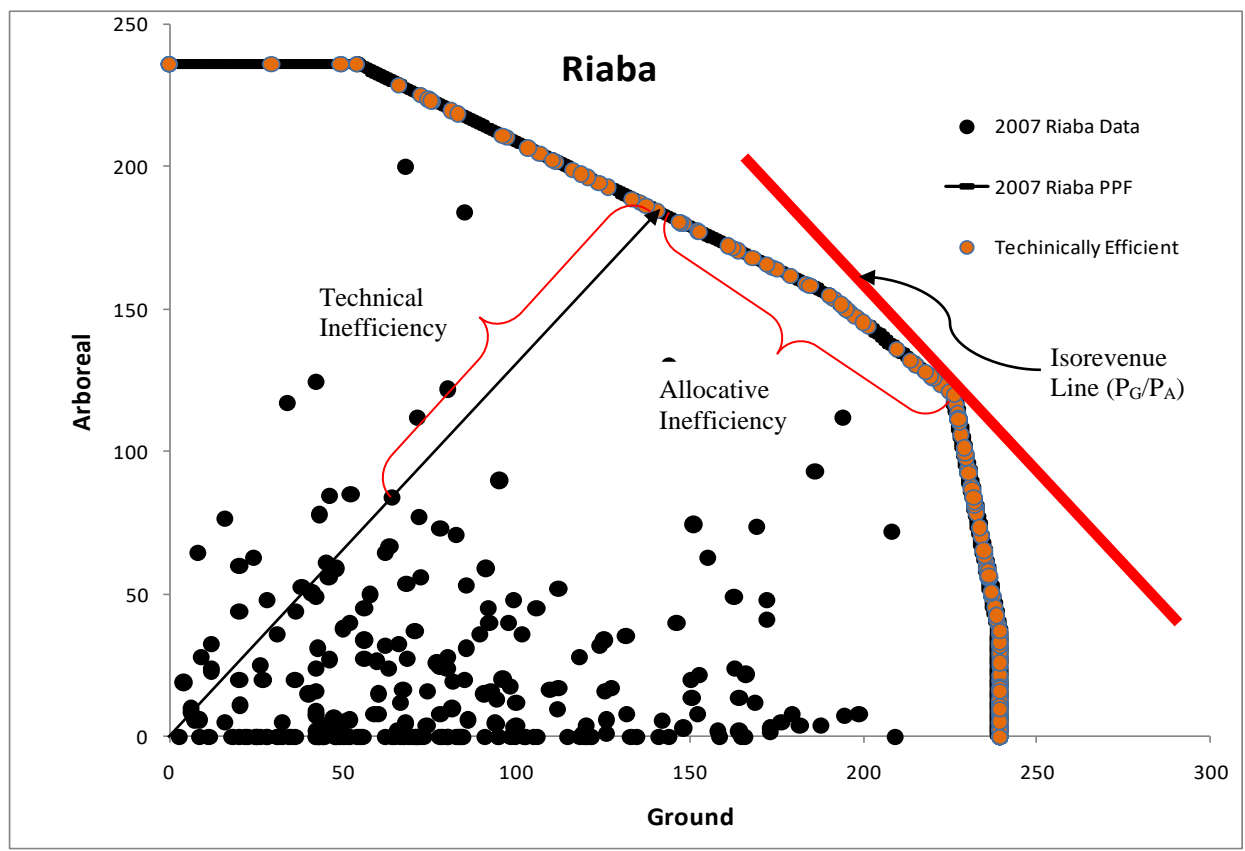

The procedure is repeated for each region (Pico Basilé, Riaba and Luba) and for each year (1999, 2001, 2003, 2005 and 2007). Table 5 and Figure 15 display the technical and allocative inefficiency for each region and year.

Table 5. Average Technical and Allocative Inefficiencies by Region and by Year.

| Location | Year | Number of Days | Average Technical Inefficiency | Standard Deviation Technical | Average Allocative Inefficiency | Standard Deviation Allocative |
|---|---|---|---|---|---|---|
| Pico Basilé | 1999 | 83 | 82 | 31.4 | 70 | 43.5 |
| Pico Basilé | 2001 | 64 | 62 | 33.8 | 140 | 19.7 |
| Pico Basilé | 2003 | 86 | 234 | 80.2 | 263 | 111.3 |
| Pico Basilé | 2005 | 209 | 132 | 44.3 | 178 | 54.3 |
| Pico Basilé | 2007 | 201 | 122 | 51.8 | 193 | 53.8 |
| Riaba | 1999 | 166 | 126 | 46.6 | 78 | 59.5 |
| Riaba | 2001 | 154 | 155 | 51.2 | 83 | 45.6 |
| Riaba | 2003 | 126 | 161 | 57.7 | 196 | 91.8 |
| Riaba | 2005 | 269 | 154 | 56.1 | 107 | 66.0 |
| Riaba | 2007 | 259 | 152 | 51.6 | 210 | 57.7 |
| Luba | 1999 | 195 | 147 | 44.8 | 95 | 18.1 |
| Luba | 2001 | 222 | 161 | 86.4 | 274 | 69.8 |
| Luba | 2003 | 219 | 199 | 62.0 | 128 | 64.8 |
| Luba | 2005 | 155 | 266 | 96.7 | 166 | 79.7 |
| Luba | 2007 | 301 | 320 | 96.6 | 151 | 133.3 |

In addition to the average inefficiency, Table 15 also reports the standard deviations. Fifteen of the observations on technical inefficiency are statistically different from zero. Only in Riaba in 1999 and Luba in 2007 is the allocative inefficiency not different from zero. These persistent inefficiencies are a result of the character of hunting as a production process and the institutional features of the bushmeat trade on the island. As skilled as a hunter may be, input and output remains stochastic with much greater variability then, say, the production of semiconductors. On the institutional side, the lack of regular transport and cold storage mitigates against allocative efficiency except by sheer chance.

Figure 15. Average Technical and Allocative Inefficiencies by Region and by Year.

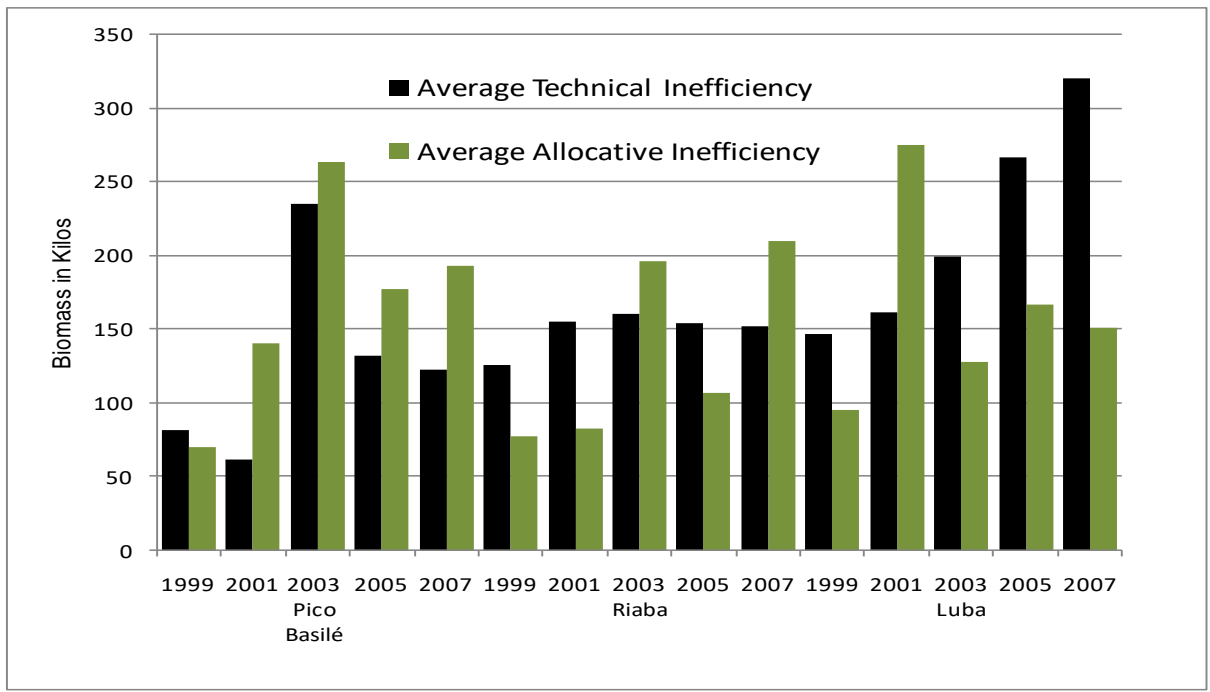

As a particular region is overexploited (Pico Basilé) or newly exploited (Luba), the post hoc changing spatial distributions of takeoff rates show evidence of a pattern of predictable inefficiency. In the north, Pico Basilé, an area of low primate density, due to excessive past takeoff rates, the technical and allocative inefficiencies rapidly increased and then tapered off. More remote southwestern Luba with its high primate populations has experienced a dramatic increase in hunting. Concomitant with the higher takeoff rates, the Luba area exhibits a persistent mounting technical inefficiency measured in biomass or revenue. At the same time, hunters appear to be more selective, targeting the more profitable species, reducing the allocative inefficiencies relative to the standard deviation.

**Sustainability**

Expanded hunting effort has consequences for wildlife that are both predictable and illuminating. Two sustainability indices (Cowlishaw (2005) and Milner-Gulland (2001)) were employed to estimate sustainable takeoff rates. Table 4 displays the excess percentage of takeoff.

Table 4. Excess Percentage of Takeoff Relative to the Maximum Sustainable Yield.

| Species of Monkey | Robinson and Redford algorithm | | US National Marine Fisheries Services algorithm | |
|---|---|---|---|---|
| | Lower Bound | Upper Bound | Lower Bound | Upper Bound |
| Red-eared monkey | 180% | 160% | 290% | 260% |
| Putty-nosed monkey | 190% | 150% | 700% | 350% |
| Crowned monkey | 170% | 120% | 320% | 270% |
| Preuss's monkey | 520% | 390% | 1,870% | 930% |
| Drill | 1,100% | 840% | 3,000% | 2,270% |
| Black colobus | 150% | 120% | 610% | 520% |
| Red colobus | 160% | 120% | 280% | 240% |

Unsurprisingly, given the large price increase of bushmeat and the overall decline of monkeys observed during surveys, the calculations indicate all takeoff rates for monkeys are well beyond sustainable levels.

**Conclusions**

This paper introduces the notions of technical and allocative efficiency to the discussion of biodiversity and sustainability. The theoretical paradigm is based on the simple notion of opportunity cost and the production possibilities frontier. The theoretical construct is applied to the harvest of bushmeat on Bioko Island, Equatorial Guinea.

There is substantial empirical evidence that the harvest of bushmeat is characterized by both technical and allocative inefficiency. For each of five years in each of three regions measured inefficiencies are significantly different from zero in 13 out of the 15 cases. This evidence leads to the inevitable conclusion that commercial bushmeat hunters are not profit maximizers in spite of their ability to target individual species. Furthermore, as shown above, the harvest rates are unsustainable.

Commercial hunting for bushmeat is the lone threat to wildlife on Bioko Island. For the most part, the largest forest mammals are taken by shotgun. Since the larger, slow-reproducing bushmeat species, especially the monkeys, are particularly susceptible to shotgun hunting, they will tend to be the next species extirpated from Bioko. Virtually all the shotgun hunting takes place within the boundaries of the two protected areas on Bioko Island and is, therefore, illegal (Ley No. 8/1988). Equatorial Guinea is a signatory to the CITES agreement and moreover has enacted laws banning the selling and hunting of endangered species (Ministerio de Pesca y Medio Ambiente, 2003) and Decree Number 72/2007, October 27, 2007, by which "the hunting, sale, consumption, and possession of monkeys and other primates in the Republic of Equatorial Guinea are strictly forbidden." Unfortunately the will by the Equatorial Guinean government to undertake enforcement of the 2007 ban is lacking.

What is unequivocal, given the estimates for current population and takeoff relative to the maximum sustainable yield is that bushmeat hunting, for the large-bodied slow-reproducing forest mammals, on Bioko is unsustainable. Given the large price increase of bushmeat and the overall decline of monkeys observed during forest surveys, the situation is not likely to change in the future.

**Recommendations**

It is possible to implement policy changes that can preserve biodiversity on Bioko Island while costs and benefits are evaluated. The banning and confiscation of shotguns on Bioko Island would stop the slaughter of monkeys by shifting the PPF inward and flattening the slope. At the same time, enforcement of existing laws prohibiting hunting in the two protected areas by trained guards/rangers would allow wildlife populations to increase. Guidelines for sustainable hunting on Bioko Island can be prepared and implemented. The two protected areas should be accurately demarked.

The scientific community can develop strategies to make conservation pay. Bioko Island provides an excellent location for study abroad educational partnerships in conservation biology and wildlife management. Hunters can be employed as guides, monitors and guards. Some local people have proven to be suitable census takers.

Lastly, the multinational corporate community must recognize that it has a stake in more than just the oil it can move out of Equatorial Guinea. Corporations can enlighten employees to not contribute to the extinction of Bioko's wildlife. Strategies include prohibiting the use of company equipment for purchasing or transporting threatened wildlife. Corporations have, and are, providing assistance, both logistical and financial, to the study and conservation of Bioko's biodiversity. Their continued support will prove to be invaluable to any future conservation strategies.

**References**

Bennett, E. T. (1833). *Cercopithecus pogonias*. Proceedings Zoological Society London, September 20, 1833, p.67. Fernando Po, West Africa. Skin without skull.

Bergl, R.A., Oates, J.F., Fotso, R. (2007). Distribution and protected area coverage of endemic taxa in West Africa's Biafran forests and highlands. Biological Conservation, Volume 134, Issue 2, January 2007, Pages 195-208.

Burton, R.E. (1863). *Wanderings in West Africa from Liverpool to Fernando Po*. Tinsley Bros., London.

Butynski, T.M. (1990). Comparative ecology of blue monkeys (*Cercopithecus mitis*) in high and low density subpopulations. *Ecological Monographs* 60:1-26.

Butynski, T.M., and Koster, S.H. (1989) The status and conservation of forests and primates on Bioko Island (Fernando Po), Equatorial Guinea: A call for research and conservation. WWF Unpublished Report, Washington D.C.

Butynski, T.M., and Koster, S.H. (1994). Distribution and conservation status of primates in Bioko Island, Equatorial Guinea. *Biodiversity and Conservation* 3:893-909.

Butynski, T. M., C. D. Schaaf, and G. W. Hearn (1997). African buffalo (*Syncerus caffer*) extirpated on Bioko Island, Equatorial Guinea. Journal of African Zoology; 111(1): 57-61.

Central Intelligence Agency (2008) World Factbook, Washington, D.C. ISSN 1553-8133 https://www.cia.gov/library/publications/the-world-factbook/geos/ek.html Last accessed 2/12/2008.

Cole, L.C. (1954) The population consequences of life history phenomena. *Quarterly Review of Biology* 29:103–137

Collel, M., Mate, C., and Fa, J.E. (1994). Hunting among Moka Bubis in Bioko: dynamics of faunal exploitation at the village level. *Biodiversity and Conservation* 3:939-950.

Cowlishaw, G. (1999). Predicting the pattern of decline of African primate diversity: an extinction debt from historical deforestation. *Conservation Biology* 13:1183-1193.

Cowlishaw, G., Mendelson, S. and Rowcliffe, J.M. (2005) Evidence for post-depletion sustainability in a mature bushmeat market. *Journal of Applied Ecology* **42**: 460-468.

Eisentraut, M. (1973). *Die Wirbeltierfauna von Fernando Poo und Westkamerun*, Bonner Zoologische Monographien, Nr. 3, Bonn.

Fa, J.E. (2000) Hunted animals in Bioko Island, West Africa: sustainability and future. In Robinson, J.G. and Bennett, E.L. (eds), *Hunting for Sustainability in Tropical Forests,* Columbia University Press, New York, pp. 168-198.

Fa, J.E., Juste B, J., Perez Del Val, J., and Castroviejo, J. (1995). Impact of market hunting on mammalian species in Equatorial Guinea. *Conservation Biology* 9:1107-1115.

Fa, J.E., Carlos A Peres, Jessica Meeuwig (2002). Bushmeat Exploitation in Tropical Forests: an Intercontinental Comparison. *Conservation Biology* 16 (1), 232–237.

Gonzalez-Kirchner, J.P. (1994). *Ecologia y Conservacion de los Primates de Guinea Ecuatorial*. Ceiba Ediciones, Cantabria.

Gonzalez-Kirchner, J.P. (1996b). Habitat preference of two lowland sympatric guenons (*Cercopithecus nictitans; C. pogonias)* on Bioko Island, Equatorial Guinea. *Folia Zoologica* 45:201-208.

Gray , J.E. (1869). Catalogue of carnivorous, pachydermatous, and edentate Mammalia in the British Museum p. 109.

Gregson (Hansen), M.E. (1993). Rural Response to Increased Demand: Crop Choice in the Midwest, 1860-1880," *Journal of Economic History* 53(2), 332-345.

Hearn, G. W. and Morra, W.A. (2000). *Census of diurnal primate groups in the Gran Caldera Volcanic de Luba, Bioko Island, Equatorial Guinea.* Unpublished report to the Government of Equatorial Guinea, Arcadia University, Glenside, PA.

Hotelling, H. (1931). "The Economics of Exhaustible Resources", *Journal of Political Economy* 39(2):137-175.

International Union for Conservation of Nature and Natural Resources (IUCN) 2007. *2007 IUCN Red List of Threatened Species*. http://www.iucnredlist.org. Downloaded on 7 August 2007.

Kingdon, J. 1997. *The Kingdom Field Guide to African Mammals.* Academic Press, San Diego. Powell, C.B., and Van Rompaey, H. 1998. Genets of the Niger Delta, Nigeria. *Small Carnivore Conservation* 19:1-7.

Kingsley, M. 1897. Travels in West Africa. Macmillan, London.

Ley Num. 8/1988 Reguladora de la Fauna Silvestre, Caza y Areas Protegidas (Law No. 8/1988 of 31 December 1988, regarding Wildlife, Hunting and Protected Areas.

Mate, C., and Colell, M. (1995). Relative Abundance of Forest Cercopithecines in Arihá, Bioko Island, Republic of Equatorial Guinea. *Folia Primatol.* 64:49-54.

Milner-Gulland, E.J., Akcakaya, H.R. (2001) Sustainability indices for exploited populations. *Trends in Ecology and Evolution* 16, 686-692.

Ministerio de Pesca y Medio Ambiente, (2003). Ley reguladora del medio ambiente en la Republica de Guinea Ecuatorial. Bata, Equatorial Guinea.

Mittermeier, R.A., Myers, N., Gil, P.R., and Mittermeier, C.G. (1999). *Hotspots: Earth's biologically richest and most endangered terrestrial ecoregions.* CEMEX, Mexico City.

Oates, J.F. (compiler) (1996). *African Primates: Status Survey and Conservation Action Plan*. Revised Edition. IUCN/SSC Primate Specialist Group, New York.

Ross, C. (1988). The intrinsic rate of natural increase and reproductive effort in primates, *J. Zool. Lond.* 214: 199–219.

Ross, C. (1998) Primate Life Histories. Evolutionary Anthropology 6, 54-63

Sundiata, I. K. (1996). From slaving to neoslavery: the Bight of Biafra and Fernando Po in the era of abolition, 1827-1930. Madison: University of Wisconsin Press.

United Nations (2003, September 12). International Human Rights Instruments, HRI/CORE/1/Add.126. Core document forming part of the reports of state parties, Equatorial Guinea.

UNHCR (2001). United Nations High Commissioner for Refugees, Equatorial Guinea: The Position of Refugees and Exiles, 2001.

Vansina, J. (1990). *Paths in the rainforests: toward a history of political tradition in equatorial Africa.* University of Wisconsin Press, Madison.

Waterhouse, G. (1838). Mammals of Fernando Po. *Proceedings of the Zoological Society of London.* 57-62.